\documentclass[cpp,a4paper,fleqn%
]{w-art}
\usepackage{times,cite,w-thm}
\theoremstyle{plain}

\theoremstyle{definition}

\usepackage[]{graphicx}

\newcommand{\mE}{\mathcal{E}}

\newcommand{\mK}{\mathcal{K}}
\newcommand{\mN}{\mathcal{N}}
\newcommand{\mD}{\mathcal{D}}
\newcommand{\ord}{\mathcal{O}}
\newcommand{\Tpe}{T_\perp}
\newcommand{\Tpa}{T_\|}
\newcommand{\vpe}{v_\perp}
\newcommand{\vpa}{v_\|}
\newcommand{\kpe}{k_\perp}

\newcommand{\Bv}{\mathbf{B}}
\newcommand{\Ev}{\mathbf{E}}
\newcommand{\bv}{\mathbf{b}}
\newcommand{\vv}{\mathbf{v}}
\newcommand{\cd}{\cdot}
\newcommand{\p}{\partial}
\newcommand{\na}{\nabla}
\newcommand{\btheta}{\bar{\theta}}
\newcommand{\phit}{\tilde{\phi}}
\newcommand{\oert}{\tilde{\omega}_{Er}}

\begin{document}
%%    The information for the title page will be placed between
%%    \begin{document} and \maketitle. The order of most entries
%%    is determined by the class file and can not be changed by
%%    rearranging them. The maketitle command follows after the
%%    abstract.
%%
%%    Most of the following commands will be completed by the publisher.
%%
%%    The copyrightyear is defined in the .clo file as the first argument
%%    of the copyrightinfo command. If the copyrightyear differs from that
%%    value it might be adjusted by the following definition:
%%
%% \renewcommand{\copyrightyear}{2007}% uncomment to change the copyrightyear.
%%
\DOIsuffix{theDOIsuffix}
%%
%% issueinfo for the header line
\Volume{46}
\Month{01}
\Year{2007}
%%
%%    First and last pagenumber of the article. If the option
%%    'autolastpage' is set (default) the second argument may be left empty.
\pagespan{1}{}
%%
%%    Dates will be filled in by the publisher. The 'reviseddate' and
%%    'dateposted' (Published online) entry may be left empty.
\Receiveddate{XXXX}
\Reviseddate{XXXX}
\Accepteddate{XXXX}
\Dateposted{XXXX}
\keywords{tokamak, impurity transport, poloidal asymmetries,
  gyrokinetic, neoclassical, peaking factor }

%% \pretitle{Editor's Choice}

%% We have a short and a long form for the title. The short form
%% (optional argument) goes into the running head.

\title[Poloidal asymmetries and impurities]{Radio frequency induced and
  neoclassical asymmetries and their effects on turbulent impurity
  transport in a tokamak}

%% Please do not enter footnotes or \inst{}-notes into the optional
%% argument of the author command. The optional argument will go into
%% the header.  If there is only one address the marker \inst{x} may be
%% omitted.

%% Information for the first author.
\author[I. Pusztai]{Istv\'{a}n Pusztai\inst{1,2,}%
  \footnote{Corresponding author\quad E-mail:~\textsf{pusztai@chalmers.se},
            Phone: +1\,617\,253\,5799,
            Fax: +1\,617\,253\,0238}}
\address[\inst{1}]{Applied Physics, Chalmers University of Technology and 
   Euratom-VR Association, SE-41296 G\"{o}teborg, Sweden}
\address[\inst{2}]{Plasma Science and Fusion Center, Massachusetts 
  Institute of Technology, Cambridge, MA 02139, USA}
%%
%%    Information for the second author
\author[M. Landreman]{Matt Landreman\inst{3}}
%%\footnote{Second author footnote.}
\address[\inst{3}]{University of Maryland, College Park, MD 20742, USA}
%%
%%    Information for the third author
\author[A. Moll\'{e}n]{Albert Moll\'{e}n\inst{1}}
\author[Ye. O. Kazakov]{Yevgen O. Kazakov\inst{4}}
\address[\inst{4}]{Laboratory for Plasma Physics, ERM/KMS, Association 
`EURATOM-Belgian State', TEC Partner, BE-1000 Brussels, Belgium} 
\author[T. F\"{u}l\"{o}p]{T\"{u}nde F\"{u}l\"{o}p\inst{1}}
%%\footnote{Third author footnote.}

%%
%%    \dedicatory{This is a dedicatory.}
\begin{abstract}
 Poloidal asymmetries in the impurity density can be generated by
 radio frequency heating in the core and by neoclassical effects in
 the edge of tokamak plasmas.  In a pedestal case study, using global
 neoclassical simulations we find that finite orbit width effects can
 generate significant poloidal variation in the electrostatic
 potential, which varies on a small radial scale. Gyrokinetic modeling
 shows that these poloidal asymmetries can be strong enough to
 significantly modify turbulent impurity peaking. In the pedestal the
 $E\times B$ drift in the radial electric field can give a larger
 contribution to the poloidal motion of impurities than that of their
 parallel streaming. Under such circumstances we find that up-down
 asymmetries can also affect impurity peaking.

\end{abstract}
%% maketitle must follow the abstract.
\maketitle                   % Produces the title.

\section{Introduction}
Close to the magnetic axis in tokamaks, where the radial gradients of
plasma parameters are too weak to destabilize microinstabilities
\cite{porkolab}, the transport of highly charged impurities is
dominated by neoclassical processes \cite{AngioniEPS}. For typical
density and temperature profiles this leads to an accumulation of
impurities in the deep core. The accumulation not only reduces the
fusion performance by diluting the plasma, but can lead to radiative
cooling and trigger magnetohydrodynamic activity \cite{mantica}.
Tungsten excursions and radiative events are observed in recent
experiments operating with tungsten plasma facing components
\cite{putterich}. The possibility of accumulation of highly charged
impurities in the core of reactor relevant tokamak plasmas, such as in
ITER, is a serious issue \cite{highzoperation}. Therefore significant
effort has been dedicated to find scenarios or active methods to
reduce core impurity accumulation.

Auxiliary radio frequency (RF) heating in the ion- and electron
cyclotron frequency ranges has been demonstrated as a possibility of
reducing impurity accumulation \cite{neu,dux,valisa}. However the
physical processes responsible for changing impurity transport are not
fully understood. In the region very close to the axis where
turbulence is not the dominant transport channel, changes in
neoclassical transport or magnetohydrodynamic activity may be
candidates \cite{putterich}. A bit further out from the center,
changes in the turbulence properties can also play an important
role. Recent studies pointed towards a possible role of poloidal
asymmetries which can be generated in RF heated plasmas
\cite{sara1,sara2,albert1,alberttem,varenna}. In plasmas lacking
strong toroidal rotation the densities of the different species are
usually assumed to be approximately flux functions. However a
temperature anisotropy of the heated species can lead to a poloidal
variation of the density of that species, which generates a poloidally
varying electrostatic field \cite{ingesson}. This field may be weak to
leave the dynamics of the main species practically unaffected, but the
response of a highly charged impurity species to these fields can be
significant. The distribution of impurities on the flux surface can
develop significant poloidal asymmetries, and their
$\mathbf{E}\times\mathbf{B}$ drift in the poloidally varying
electrostatic field can compete with the magnetic drifts
\cite{albert1}, which has important consequences on the turbulent
impurity transport.

 The local value of the steady state radial logarithmic impurity
 density gradient (referred to as the \emph{impurity peaking factor}),
 is determined by the local turbulent transport. On the one hand,
 impurity transport in the plasma edge provides a boundary condition
 for the impurity density profile, thus affecting impurity
 accumulation in the core. On the other hand, a relatively high
 impurity concentration may be necessary in the edge for reactor
 relevant devices to reduce localized heat loads on plasma facing
 components through radiation. It is therefore of interest to know the
 impurity transport in the edge. It would be an unreasonably large
 undertaking to give a complete description of the edge impurity
 transport in one paper given the complex nature of the
 problem. Instead, we will consider certain aspects of the problem
 related to poloidal asymmetries.

There are experimental observations of strong poloidal in-out
asymmetries in the impurity density in the edge plasma
\cite{marr,churchill,AUGasymmetry}, especially in the strong gradient
regions of edge transport barriers (henceforth, we refer to the
inboard/outboard impurity accumulation as in-out/out-in density
asymmetry). In the same time, neoclassical theory also predicts
poloidal impurity density
asymmetries\cite{tunde1,tunde2,landreman}. It has to be noted that the
poloidal asymmetries experimentally observed in the pedestal can be
several times larger \cite{churchill} than what the theory described
in Refs.~\cite{tunde1,tunde2,landreman} predicts. This may partly be
due to the simplified magnetic geometries considered, but more likely
the reason is that kinetic effects associated with finite ion orbit
width \cite{peterped} in the pedestal are not accounted for.

In the present paper we first give an overview of research done on
turbulent impurity transport in the presence of poloidal asymmetries
generated by ion cyclotron resonance heating (ICRH). Then we consider
the effect of edge asymmetries due to neoclassical effects. We use a
global numerical tool for neoclassical calculations in the pedestal
PERFECT (Pedestal \& Edge Radially global Fokker-Planck Evaluation of
Collisional Transport) \cite{perfect}, and a combination of
gyrokinetic simulations and semi-analytical modeling to study the
turbulent transport in the presence of poloidal asymmetries and strong
radial electric fields.

The subsequent sections are organized as follows. In
Sec.~\ref{sec:asym} different mechanisms leading to poloidal impurity
asymmetries in the core and edge are reviewed. Then, the gyrokinetic
modeling of turbulent impurity transport in the presence of
asymmetries is described in Sec.~\ref{modeling}. Finally, simulation
results are presented and interpreted in Sec.~\ref{simulations},
before we draw our conclusions in Sec.~\ref{discussion}.

\section{Asymmetries in the core and in the edge} 
\label{sec:asym}
In numerical ICRH simulations the non-Maxwellian part of the
distribution function of the heated ion species (denoted by subscript
$m$, as minority) is often found to exhibit velocity space structures
that can be well approximated by the ansatz \cite{dendy}
$f_m\propto\exp[-\mu B_c/T_\perp-|\mK-\mu B_c|/T_\|]$, where
$\mK=m_mv^2/2$, $\mu=m_mv_\perp^2/(2B)$, and $T_\perp$ and $T_\|$ are
radially varying parameters representing the perpendicular and
parallel temperatures, respectively (note that the temperature
anisotropy $\alpha_T=\Tpe/\Tpa>1$ due to the ICRH heating).
Furthermore, the magnetic field strength $B$ is equal to $B_c$ where
the ICRH absorption is the largest on the flux surface.  The density
moment of this distribution function yields a poloidally varying
minority density with a maximum close to the poloidal angle $\theta_0$
where $B(r,\theta_0)=B_c$, which reflects that the heated particles
tend to get trapped due to the heating with their banana tips close to
the ICRH resonance location. In particular, when the $B=B_c$ surface
is tangential to the studied flux surface from the low magnetic field
side (this is the situation considered unless stated otherwise), the
density of the heated minority species has a maximum at the outboard
mid-plane ($\theta_0=0$). The corresponding poloidally varying
electrostatic potential $\phi_E$, that sets up to sustain
quasineutrality, has a minimum on the inboard side. Consequently the
impurities accumulate in that region.

In a large aspect ratio circular cross section plasma the poloidal
variation of the impurity density is approximately sinusoidal
$n_z(r,\theta)=n_{z0}(r)\exp\left[-Ze\phi_E(r,\theta)/T_z(r)\right]
\approx n_{z0}(r)\exp[K \cos(\theta-\bar{\theta})],$ where $n_{z0}$
(sometimes referred to as the pseudo density) is a flux function, $Z$
is the impurity charge number, and $T_a$ denotes the temperature of
species $a$. Furthermore $\theta=\btheta$ corresponds to the poloidal
location of the maximum of the impurity density.  The asymmetry
strength $K$ can be approximated as $K=Z k (n_{m0}/n_e)/[(n_i/n_e)
  +(T_z/T_e)+(n_{z0}Z^2/n_e)]$, with $k=[\epsilon
  b_c(\alpha_T-1)]/[b_c+\alpha_T(1-b_c)]$ \cite{kazakov1}. Here,
$n_{m0}$ is the flux surface average of the minority density,
$\epsilon=r/R_0$ with $R_0$ the major radius of the centroid of the
flux surface, $b_c=B_c/B_0$ with $B_0=B(R_0)$. We assume $T_z=T_i$ and
that the poloidal variations of the main ion and electron densities --
$n_i$ and $n_e$, respectively -- are negligible.

In the edge, where the plasma parameter profiles may be sharp due to
transport barriers there can be various effects which give rise to
poloidal asymmetries in the impurity density. In the parallel impurity
momentum equation normally the electric field and the pressure
gradient terms balance,
$T_i\nabla_\|n_z=-e_zn_z\nabla_\|\tilde{\phi}$, leading to a Boltzmann
response of the impurities (assuming $T_z=T_i$ to be a flux
function). This means that even if the impurities are not too
collisional, there is a possibility to develop poloidal impurity
density asymmetries, since parallel electric fields can be present due
to poloidal asymmetries in the main ion density. These asymmetries
naturally arise when the main ions are in the plateau or
Pfirsch-Schl\"{u}ter regime of collisionality even in the presence of
weaker gradients. The form of this density variation is
$\propto\sin\theta$ \cite{landreman}. When the gradients and the
collisionality are sufficiently high, so that the impurity-ion
friction becomes comparable to the parallel pressure gradient term in
the parallel momentum balance equation, that can also lead to poloidal
impurity density variations. In this case, if the main ions are either
in the banana or in the Pfirsch-Schl\"{u}ter regimes, the impurities
are pushed to the inboard side \cite{tunde1,tunde2}, whereas in the
plateau regime, there can be in-out or out-in impurity density
asymmetry depending on the ratio of the density and temperature scale
lengths, the strength of the impurity ion friction, and the impurity
concentration \cite{landreman}. Finite Mach number effects can also
modify the poloidal density variation and the neoclassical transport
of impurities \cite{tunde1}.

Neoclassical calculations and simulation codes normally assume that
the orbit width of particles (which, in general, is comparable to
their poloidal Larmor radius $\rho_{\theta a}=(B/B_{\theta})\rho_a$,
where $B_{\theta}$ is the poloidal magnetic field) is much smaller
than the radial scale length of plasma parameters. In this case a
local analysis (i.e. considering a single flux surface) is
sufficient. In very sharp pedestals this assumption may break down,
first for the main ions (and low charge number impurities). If the ion
temperature scale length $L_{Ti}=-[\p_r(\ln T_i)]^{-1}$ becomes
comparable with $\rho_{\theta i}$, then the distribution function may
not remain close to a Maxwellian anymore, which requires a global,
nonlinear description. However, if only the  electron density and
temperature scale lengths ($L_{ne}=-[\p_r(\ln n_e)]^{-1}$ and
$L_{Te}=-[\p_r(\ln T_e)]^{-1}$) are comparable to $\rho_{\theta i}$,
although a global description is still necessary, the distribution
function can be expanded about a Maxwellian, and the linear
drift-kinetic equation may be employed \cite{perfect}. This limit,
apart from being convenient for theoretical studies, has physical
relevance as well; the ion temperature scale length often appears to
be much larger than the density scale length in pedestals of DIII-D
\cite{groebner1,groebner2}.  Furthermore, in very large aspect ratio
$\sqrt{\epsilon}\ll 1$, the scale separation between the ion orbit
width $\sim\rho_{\theta i}\sqrt{\epsilon}$ and $\rho_{\theta i}$
allows for simplifications making an analytical treatment possible
\cite{grishaped,istvanped,peterped}. In pedestals with realistic
aspect ratio further poloidal asymmetries in the main ion flow
\cite{perfect} and density can arise. To calculate these asymmetries
one needs to resort to numerical simulations.

\section{Gyrokinetic modeling}
\label{modeling}
We consider the peaking of a highly charged trace impurity species
sustained by the linear flux of a single unstable mode. Specifically,
to determine the impurity peaking factor in steady state we require
the flux surface average of the radial linear impurity flux to vanish
$\Gamma_z\equiv-{\rm Im} \langle(k_y/B)\int d^3v J_0(z_z)g_z\phi^\ast
\rangle$. Here $\langle\cdot\rangle$ denotes the flux surface average,
$k_y=nq/r$ is the binormal wave number with $n$ being the toroidal
mode number and $q$ the safety factor, and $\phi^\ast$ is the complex
conjugate of the perturbed electrostatic potential. The impurity
finite Larmor radius (FLR) parameter $z_z$, appearing in the argument
of the Bessel function of the first kind $J_0$, is defined as
$z_z=\kpe\vpe/\Omega_z$ with $\kpe=k_y(1+s^2\theta^2)^{1/2}$ the
perpendicular wave number, $s$ the magnetic shear and
$\Omega_a=e_aB/m_a$ the cyclotron frequency of species $a$, where
$e_a$ and $m_a$ are the species charge and mass.

We formulate the problem in the laboratory frame, which is convenient
for a pedestal with subsonic ion rotation. We consider the situation
when the large diamagnetic and $\Ev\times\Bv$ rotation speeds nearly
cancel for the main ions, and the impurities are collisionally coupled
to them thereby having a similar flow speed and temperature. The
non-adiabatic part of the perturbed impurity distribution function
$g_z$ is governed by the linear gyrokinetic (GK) equation. Considering
a circular cross section large aspect ratio plasma the GK equation
reads
\begin{equation}
(\vpa+u)\bv\cd\na\theta \p_\theta
g_z-i(\omega-\omega_{Dz}-\omega_E-\tilde{\omega}_{Er})g_z-C_z^{(l)}[g_z]= -i
(e_z\phi/T_z)f_{z0}(\omega-\omega_{\ast z}^T-\tilde{\omega}_{Er})J_0(z_z),
\label{gkeq}
\end{equation}
where $\bv\cd\na\theta= 1/(qR_0)$, with $\bv=\Bv/B$ the unit vector
along the magnetic field $\Bv$, and $C_z^{(l)}$ is the linearized
impurity collision operator. The parallel gradient is to be taken at
fixed total unperturbed energy $\mE=m_zv^2/2+e_z\phi_E$ and magnetic
moment $\mu=m_z\vpe^2/(2B)$. We split
$\phi_E=\Phi(r)+\phit(r,\theta)$, where $\langle\phit\rangle=0$. The
contribution of the $\Ev\times\Bv$ drift to the poloidal motion is
represented by $u\approx B_\theta^{-1} \p_r \Phi$, the Doppler shifted
frequency is $\omega=\underline{\omega}-\omega_{Er}$ with
$\underline{\omega}$ the laboratory frame frequency and
$\omega_{Er}=(k_y/B)\p_r \Phi$, while $\oert=(k_y/B)\p_r\phit$. The
radial drift due to the poloidal electric field is
$\omega_E=-k_ys\theta(\p_\theta \phit)/(Br)$.  The magnetic and
diamagnetic drift frequencies are defined as
$\omega_{Dz}=-(k_y/\Omega_z R) (\vpa^2+\vpe^2/2)(\cos \theta
+s\theta\sin\theta)$ and $\omega_{\ast
  z}^T=\omega_{az}\{(a/L_{nz})+[m_zv^2/(2T_z)-3/2](a/L_{Tz})\}$, where
$\omega_{az}=-k_y T_z/(e_zB a)$ with $a$ the plasma minor radius, and
$1/L_{nz}=-\partial_r(\ln n_{z})$ defines the impurity density-, and
$1/L_{Tz}=-\partial_r(\ln T_z)$ the impurity temperature scale
lengths; note that $1/L_{nz}$ is not a flux function.  The lowest
order distribution function is a Maxwellian
$f_{z0}(r,\mE)=n_{z0}(m_z/2\pi T_z)^{3/2}
\exp(-\mE/T_z)=n_{z}(m_z/2\pi T_z)^{3/2} \exp(-m_zv^2/(2T_z))$, where
$n_{z0}=n_z\exp(e_z\phi_E/T_z)$ is a flux function. In the right hand
side of the gyrokinetic equation $\omega-\omega_{\ast
  z}^T-\tilde{\omega}_{Er}$ stems from
$\underline{\omega}-\tilde{\omega}_{\ast z}^T$, where
$\tilde{\omega}_{\ast z}^T=-\omega_{az} a[\p_r(\ln
  n_{z0})+(\mE/T_z-3/2)\p_r(\ln T_{z})]$.

We solve (\ref{gkeq}) perturbatively keeping $\omega$ and the
impurity-impurity collisions in $0^{\rm th}$ order, the $\vpa$ and $u$
terms in the $1^{\rm st}$ order, and the $\omega_{Dz}$, $\omega_{E}$,
$\oert$, $\omega_{\ast z}^T$ and $J_0(z_z)-1$ terms to $2^{\rm nd}$
order. This reduces to an expansion in $Z^{-1/2}$ in the plasma core,
consistent with $e_z\p_\theta(\phit)/T_z\sim 1$ (see \cite{varenna}
for details), but in the pedestal where radial scale lengths of $n_Z$
and $e_z\tilde{\phi}/T_z$ are comparable to $\rho_{\theta i}$ it is an
ad-hoc ordering to be justified \emph{a posteriori} by numerical
results. The largest term in the impurity-ion collision operator is a
friction term of the form $\propto \nabla_v f_z(\vv) \cdot \int d^3 v'
f_i(\vv') \vv'/v'^3$ that includes the perturbed ion distribution
$f_i$. We assume that this term acts to make the impurity flow speed
sufficiently close to the main ion flow speed so that $C^{(l)}_{zi}$
only needs to be included in the perturbative scheme at $2^{\rm nd}$
order. When ordering $C^{(l)}_{zi}$ this way and using the
conservation properties of $C^{(l)}_{zz}$, collisions do not affect
the turbulent impurity flux to the accuracy of the model.

Since $n_z$ varies over the flux surface we define an effective
impurity peaking factor as $ a/L_{nz}^0=a/L_{nz}^\ast$ when
$\Gamma_z=0$, where $a/L_{nz}^\ast=\langle a/L_{nz}\rangle_\phi$ with
the weighted flux surface average
$\langle\dots\rangle_\phi=\langle\dots \mN |\phi|^2\rangle/\langle\mN
|\phi|^2\rangle$ and $\mN(\theta)=\exp(-e_z\phit/T_z)$. Assuming
$Ze\phit/T_z=-K \cos(\theta-\btheta)$ and neglecting $\ord(\epsilon)$
corrections we find
\begin{align}
\frac{a}{L_{nz}^0}= &2 \frac{a}{R_0}\langle \mathcal{D} \rangle_\phi +
\frac{a}{r}s K\langle\theta \sin(\theta-\btheta) \rangle_\phi\label{peak1}\\
-&\left(\frac{m_i}{m_z}+2U_i^2\right) \frac{2 Z a v_i}{q^2
  R_0^2k_y\rho_i} \frac{\omega_r}{|\omega|^2} \left\langle
\frac{|\partial_\theta \phi|^2}{|\phi|^{2}} \right\rangle_\phi+\frac{2
  U_i Z a}{q R_0 k_y\rho_i{\rm Im}[\omega^{-1}]} \left\langle {\rm
  Im}\left[\frac{i}{\omega} \frac{\p_\theta \phi}{\phi}\right]\nonumber
\right\rangle_\phi,
\end{align}
where $\mD=\cos\theta+s\theta\sin \theta$, the real part of $\omega$
is $\omega_r$, $U_i=u/v_i$, and $\rho_i=v_im_i/eB_0$ with
$v_i^2=2T_i/m_i$.  The first and second terms of (\ref{peak1})
represent contributions from the magnetic drifts and the
$\Ev\times\Bv$ drift in the poloidal electric field, respectively. The
terms in the second line of (\ref{peak1}) are contributions from the
poloidal motion of the particle due to parallel streaming and the
$\Ev\times\Bv$ drift in the radial electric field. The $U_i^2$ and
$U_i$ terms can be important in the pedestal where $U_i$ can reach
$\mathcal{O}(1)$ values. Note that if $\phi$ is symmetric around
$\theta=0$ the last term of Eq.~(\ref{peak1}) is non-zero only if
$\phit$ is not even in $\theta$ (i.e. there is a finite up-down
asymmetry). Outside the pedestal ($U_i\ll 1$) up-down asymmetric potentials have
negligibly small effect on the turbulent impurity transport as
demonstrated in \cite{alberttem}.

\section{Simulation results and interpretation}
\label{simulations}
In this section we demonstrate the importance of neoclassical
asymmetries on turbulent impurity transport in an edge transport
barrier through numerical modeling.

The modeling is done in two steps. To calculate the neoclassical
poloidally varying electrostatic potential $\tilde{\phi}$, first we
use the code PERFECT \cite{perfect}, which is a radially global,
$\delta f$, continuum neoclassical solver. PERFECT allows the electron
density and temperature scale lengths to be comparably small to the
poloidal ion Larmor radius, $\rho_{\theta i}$, while requiring the ion
temperature to vary slowly on that scale. The potential $\tilde{\phi}$
is then used as an input to the modeling of the turbulent impurity
transport.

\begin{figure}[h]
\begin{center}
\includegraphics[width=0.3275\linewidth]{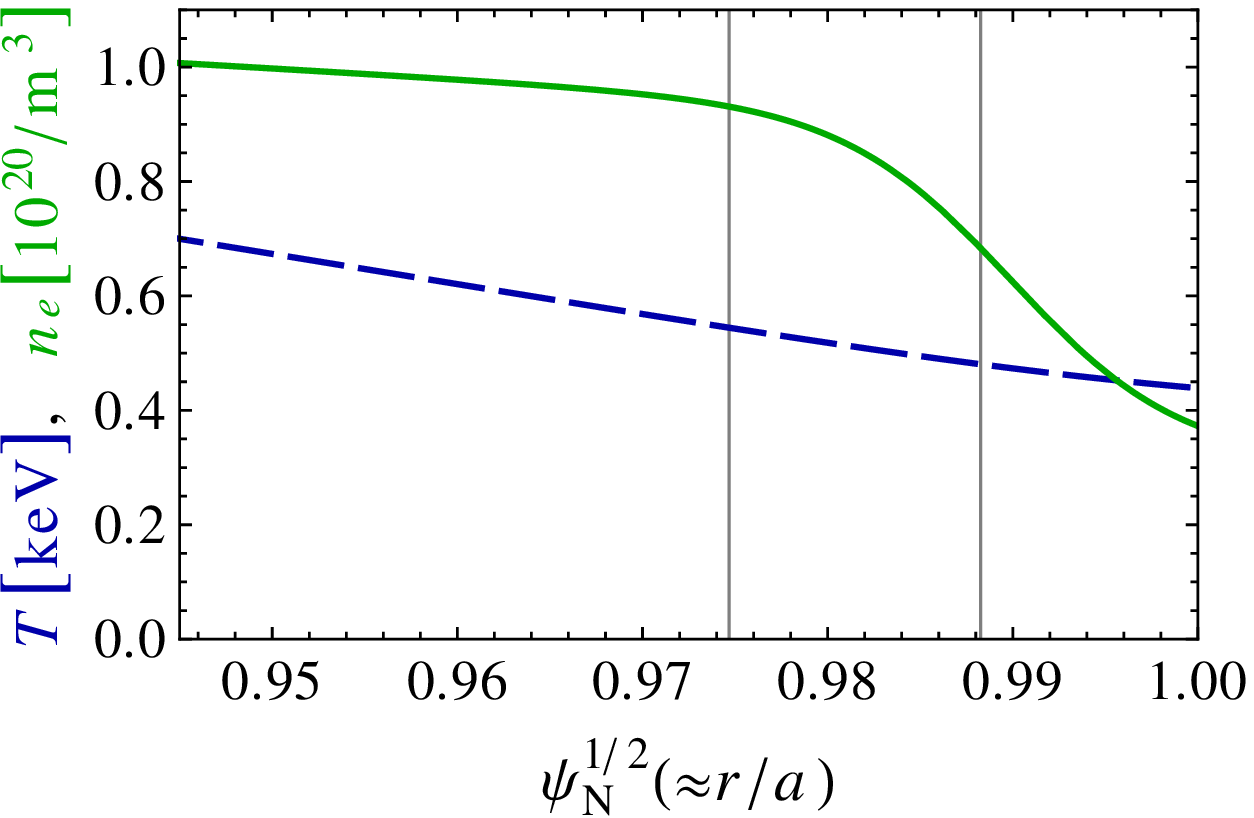}
\put(-20,30){$\rm a)$}
\put(127,30){$\rm b)$}
\put(275,30){$\rm c)$}
\includegraphics[width=0.3275\linewidth]{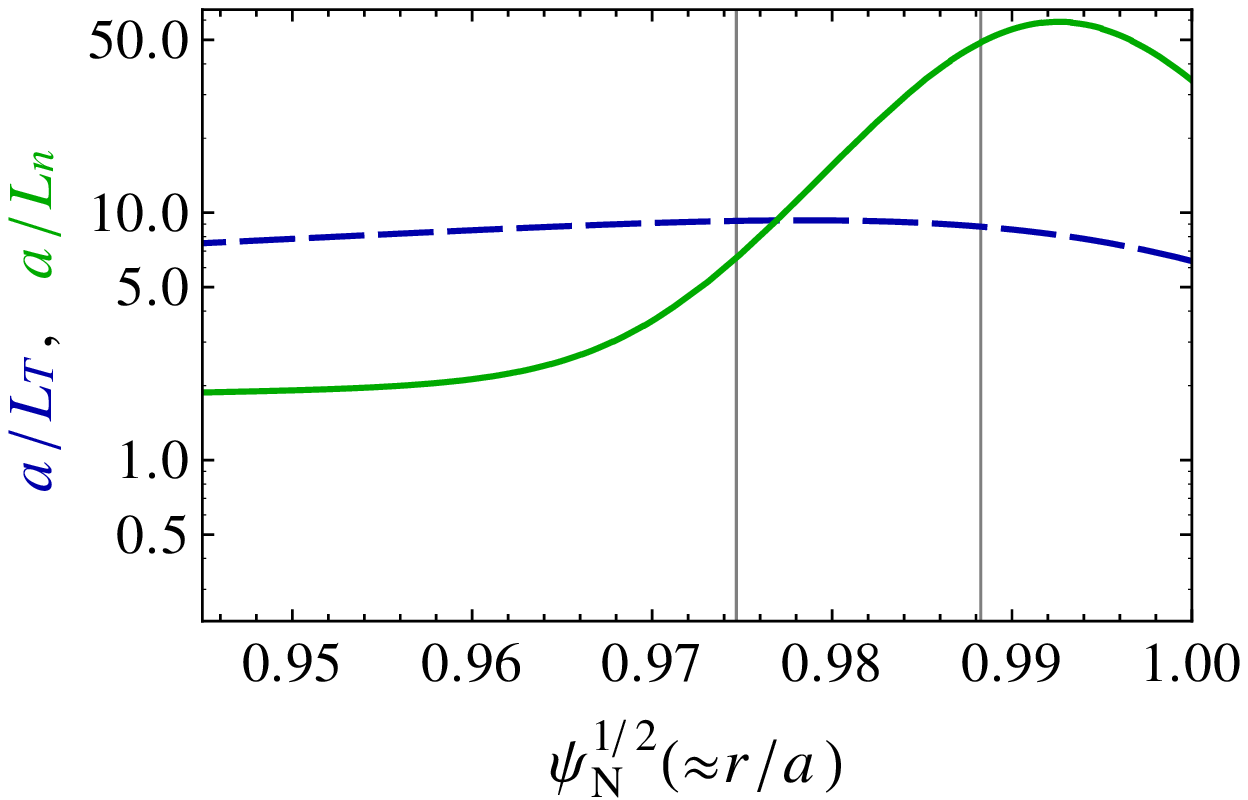}
\includegraphics[width=0.3275\linewidth]{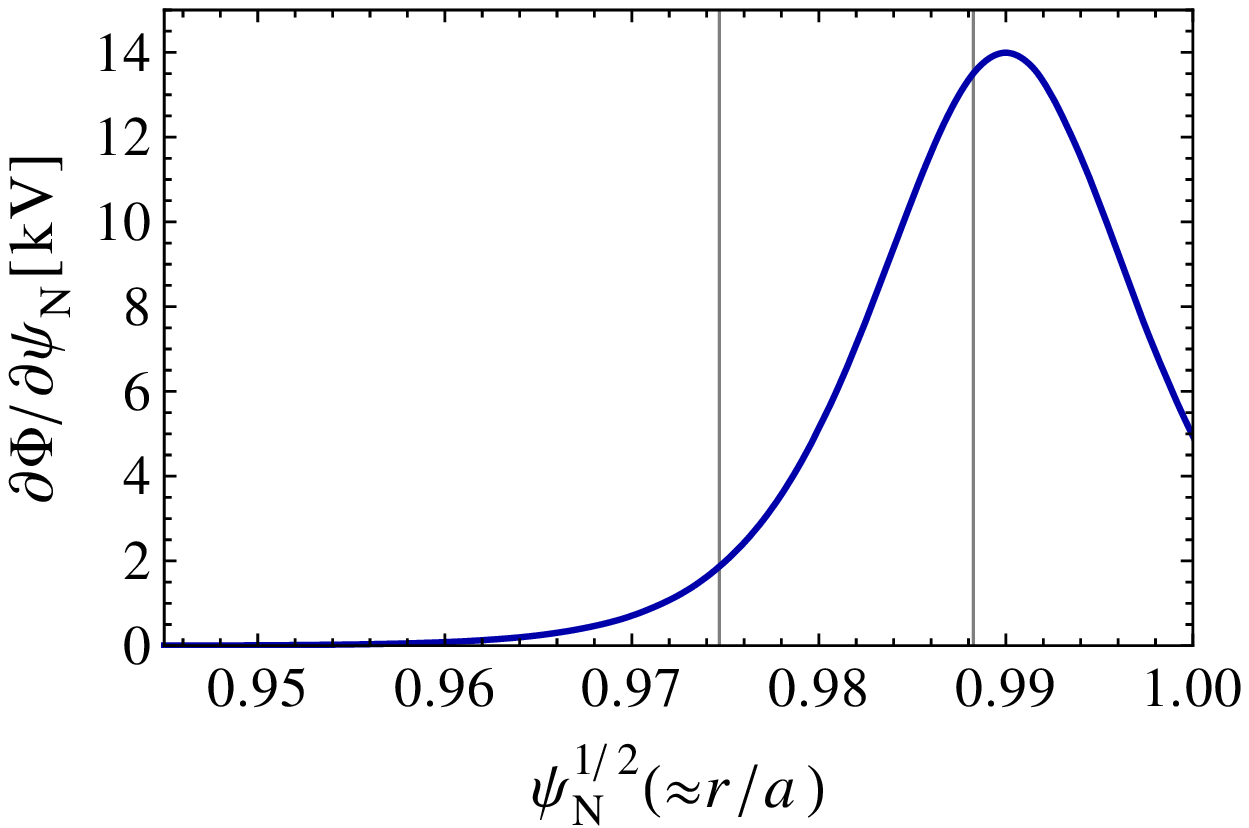}
\caption{Variation of plasma parameters across the simulation domain,
  used as inputs for the PERFECT simulation. The following quantities
  are depicted: a) temperature in keV [dashed curve], electron density
  in $10^{20}\rm{m}^{-3}$ [solid curve], b) $a/L_T$ [dashed], $a/L_n$
  [solid], c) radial variation of the electrostatic potential in kV.}
\label{inprofiles}
\end{center}
\end{figure}

We consider plasma parameter profiles that resemble experimental
profiles in H-mode pedestals of the DIII-D tokamak, although they do
not represent any specific experiment. The neoclassical simulations
aim to show that it is plausible for finite-orbit-width neoclassical
physics to generate in-out and/or out-in asymmetries; indeed these
asymmetries are observed for a wide range of input profiles in
PERFECT.  Figure~\ref{inprofiles} shows the variation of the most
important plasma parameters across the simulation domain in the
PERFECT simulation. The ion and electron temperatures, which are
assumed to be equal, vary slowly over the simulation domain, as shown
by the dashed curve in Fig.~\ref{inprofiles}a. However, the electron
density has a significant drop around $\psi_N^{1/2}=0.99$ (solid
curve). Accordingly, $a/L_n(\equiv a/L_{ne}\approx a/L_{ni})$ greatly
exceeds $a/L_T(\equiv a/L_{Ti}=a/L_{Te})$ in this region, as shown in
Fig.~\ref{inprofiles}b. The maximum value of $\rho_{\theta i}/L_n$ and
$\rho_{\theta i}/L_T$ in the domain are $0.92$ and $0.16$,
respectively. The impurities are assumed to be trace in the sense that
they do not affect quasineutrality, hence the electron and ion density
profiles are approximately the same. In particular, the PERFECT
simulation was performed for a pure plasma. The simulation uses a
Miller model equilibrium and the variation of the magnetic geometry
parameters across the domain is neglected. The aspect ratio is
$R/a=3.17$, the radial variation of the Shafranov shift is $\p_r
R_0=-0.354$, the Miller parameters \cite{miller} for elongation are
$\kappa=1.66$ and $s_\kappa\equiv (r\p_r \kappa)/\kappa=0.7$, and
those for the triangularity are $\delta=0.416$ and
$s_\delta\equiv(r\p_r\delta)/(1-\delta^2)^{1/2}=1.37$, finally, the
safety factor is $q=4$, and the magnetic shear is unimportant for the
neoclassical simulation. The radial variation of the potential $\Phi$
corresponding to the radial electric field is not calculated in
PERFECT but rather it needs to be provided as an input for the
calculation; see Fig.~\ref{inprofiles}c.

The converged resolution parameters of the simulation are the
following. The number of poloidal modes is $N_\theta=25$, the number
of Legendre polynomials used in the Rosenbluth potentials is $N_L=4$,
the number of grid points in $v$, $\xi=\vpa/v$ and normalized poloidal
flux $\psi_N$ (the radial coordinate) are $N_v=6$, $N_\xi=13$ and
$N_\psi=55$, respectively. The domain size in speed is $v_{\max}=5
v_i$. The simulation considered the dynamics of a single ion species,
deuterium.

\begin{figure}[h]
\begin{center}
\includegraphics[width=0.35\linewidth]{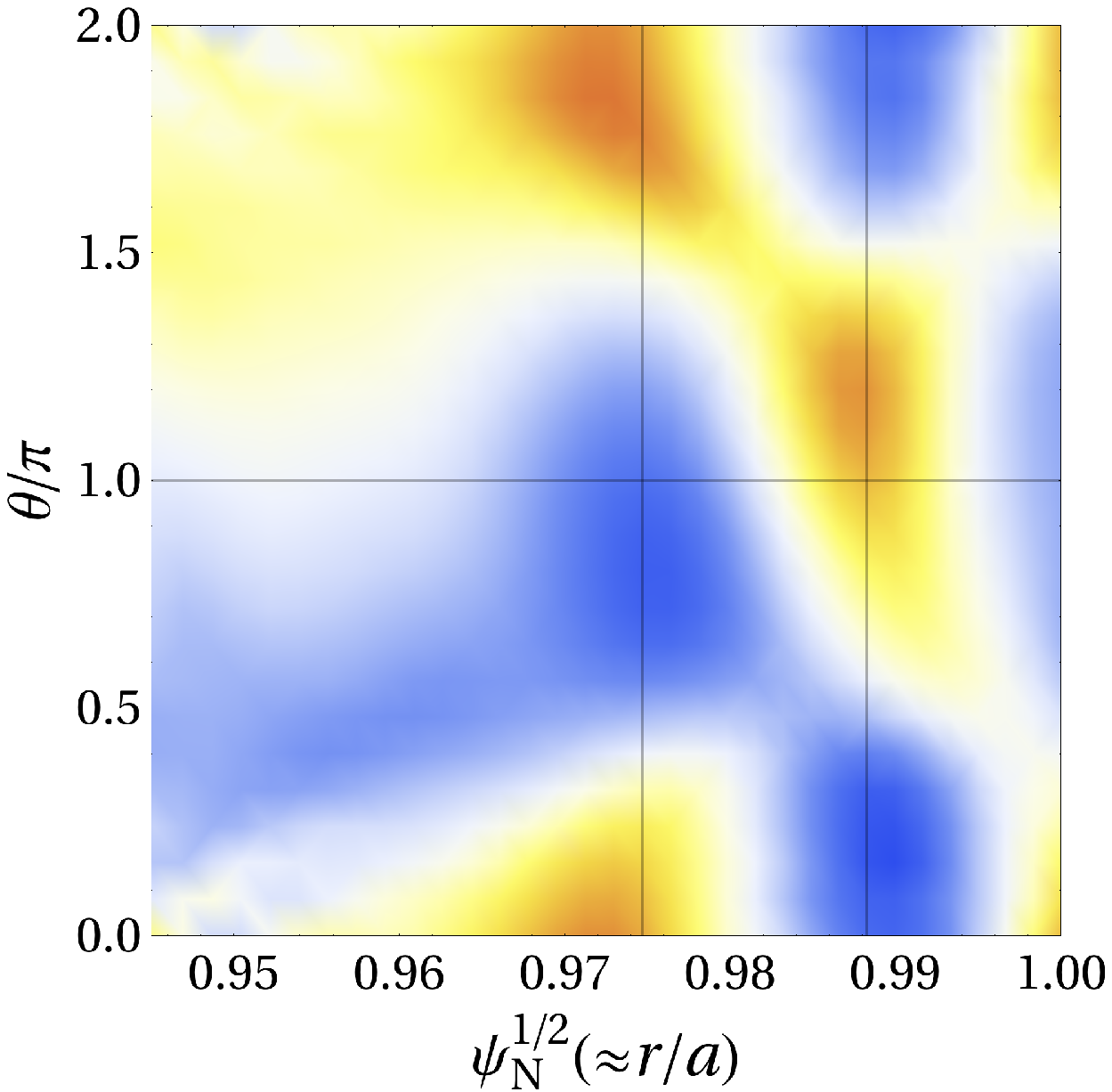}
\put(-125,30){$\rm a)$}
\put(150,30){$\rm b)$}
\includegraphics[width=0.4\linewidth]{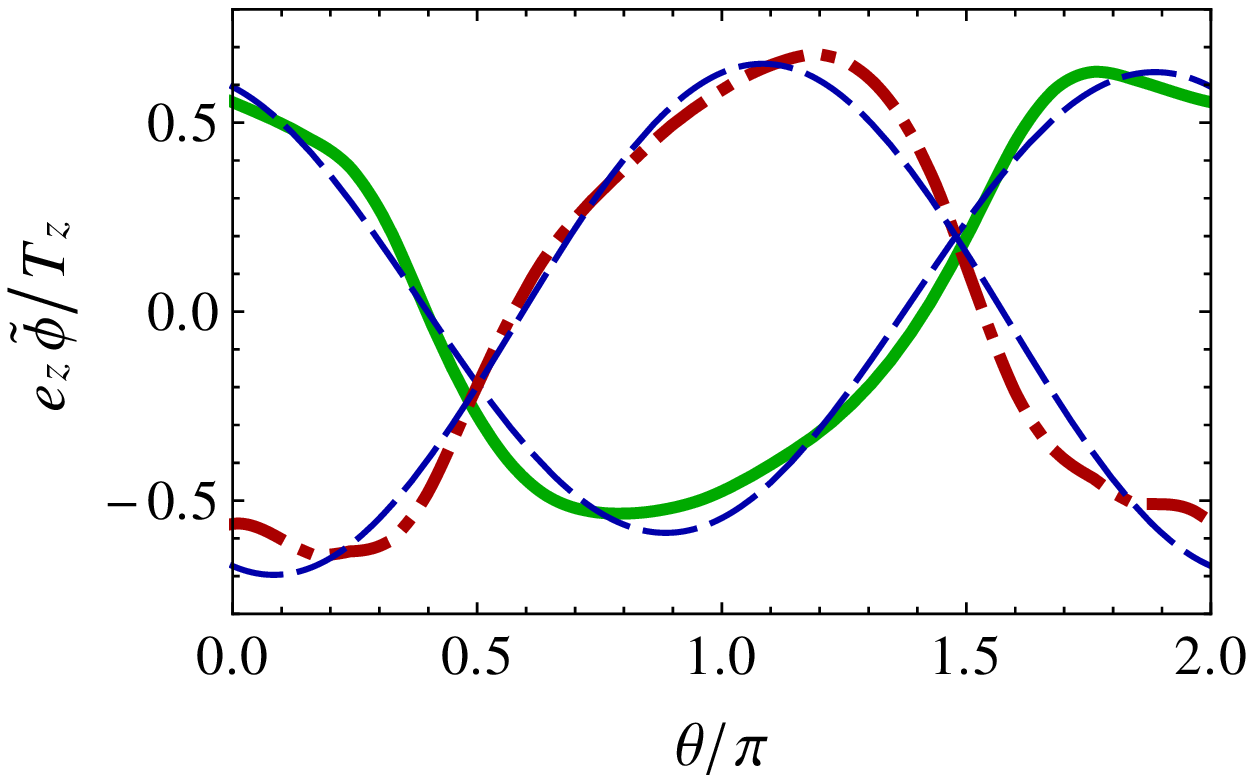}
\caption{ a) Perturbed electrostatic potential, $\tilde{\phi}$, as a
  function of radius and poloidal angle (output of the PERFECT
  simulation): the largest positive (red) value is $15.2\mathrm{V}$,
  the highest negative (blue) value is $-11.5\mathrm{V}$. The vertical
  bars mark the two studied radial locations, $\psi_N^{1/2}=0.975$ and
  $\psi_N^{1/2}=0.988$.  b) The poloidal variation of
  $e_z\tilde{\phi}/T_z$ at $\psi_N^{1/2}=0.975$ (solid line) and at
  $\psi_N^{1/2}=0.988$ (dash-dotted line), together with corresponding
  sinusoidal fits (thin dashed lines).  }
\label{potentialpert}
\end{center}
\end{figure}

Out of the several outputs of the PERFECT simulation we only need the
poloidally varying electrostatic potential $\tilde{\phi}$ that is
generated by the perturbed ion density. This quantity is shown in
Fig.~\ref{potentialpert}a as a function of $\sqrt{\psi_N}$ and
$\theta$. The red (blue) tones correspond to positive (negative)
values of the potential. The most conspicuous feature of
$\tilde{\phi}$ is that the poloidal location of the minimum of the
potential can vary rapidly radially. On the inner boundary of the
simulation domain there is a moderate up-down asymmetry, which
transits to a stronger in-out asymmetry. Then around the location of
the pedestal top the polarity flips from in-out to an out-in asymmetry
which, just inside the last closed flux surface (LCFS) flips back to
an in-out asymmetry.

Since up-down asymmetries in $\tilde{\phi}(\theta)$ are shown to have
only a very weak effect on turbulent impurity transport if $U_i\ll 1$
we concentrate on areas with in-out or out-in asymmetries. It might be
worth emphasizing that conventional neoclassical theory based on the
assumption of $L_T\sim L_n\gg \rho_{\theta i}$ can provide only
up-down potential asymmetry (as long as the flux surface is not
strongly up-down asymmetric). Accordingly, an advanced theory or
simulation code is necessary to compute $\tilde{\phi}(\theta)$.  We
choose two radial locations for further studies from the confined
region ($\psi_N<1$) where the asymmetries are the strongest. These
locations, $\psi_N^{1/2}=0.975$ and $\psi_N^{1/2}=0.988$, are marked
with vertical bars in Fig.~\ref{potentialpert}a, and in
Fig.~\ref{inprofiles} as well. In the following we will calculate the
effective impurity peaking factors at these two locations using the
model presented in Section~\ref{modeling}. As an impurity we consider
fully ionized nickel ($Z=28$). For this charge number, the poloidal
variation of $e_z\tilde{\phi}/T_z$ is shown in
Fig.~\ref{potentialpert}b for the two radial locations. We approximate
the potential variation with the sinusoidal model given above
Eq.~(\ref{peak1}); the corresponding fitted parameters for
$\psi_N^{1/2}=0.975$ are $K=0.610$ and $\btheta/\pi=0.885$, and for
$\psi_N^{1/2}=0.988$ are $K=0.676$ and $\btheta/\pi=0.084$ (the fits
are represented by the dashed lines in Fig.~\ref{potentialpert}b).

Toroidal rotation and rotation shear can modify heavy impurity
transport significantly, as demonstrated in Ref.~\cite{angioni} and
references therein. Assuming the parallel ion and impurity flow speeds
to be comparable, the square of the impurity Mach number is
$M_z^2\approx 0.16$ on flux surface average at these locations. This
may be non-negligible, but it is expected to have a smaller effect on
the impurity peaking factor than that of the poloidally varying
electrostatic potential.  For simplicity the radial coordinate $r/a$
is taken to be equal to $\psi_N^{1/2}$.

The local plasma parameters used in the simulations at the two radial
locations are the following. At $r/a=0.975$, $T_i/T_e=1$,
$a/L_{ne}=a/L_{ni}=6.59$, $a/L_{Te}=a/L_{Ti}=9.27$, and the
electron-ion collision frequency is $\nu_{ei}[c_s/a]=1.72$, where
$c_s=(T_e/m_i)^{1/2}$ is the ion sound speed.  At $r/a=0.988$,
$T_i/T_e=1$, $a/L_{ne}=a/L_{ni}=48.7$, $a/L_{Te}=a/L_{Ti}=8.78$, and
$\nu_{ei}[c_s/a]=1.62$. The same magnetic geometry parameters are used
as in the neoclassical simulation, and additionally we set the
magnetic shear $s=5$ in both radial locations.

We use local, electrostatic gyrokinetic simulations with the GYRO code
\cite{gyro} to calculate the mode frequency $\omega=\omega_r+i\gamma$
and the eigenfunction $\phi$. In the GYRO simulations we consider only
electrons and ions, assuming a trace impurity, and use $\omega$ and
$\phi$ as inputs for the model represented by
Equation~(\ref{peak1}). Since the main species are very weakly
affected by the poloidal variation of the neoclassical potential, it
does not pose a problem that GYRO assumes poloidally symmetric
distribution functions. To our knowledge, presently there exists no
gyrokinetic simulation code that consistently accounts for finite
orbit width effects arising due to $U_i\sim 1$, which is also the case
in GYRO; accordingly such effects are not taken into account in the
gyrokinetic simulations, although $U_i\sim 1$ in the pedestal. The
simulations use GK ions and drift kinetic electrons with
$(m_i/m_e)^{1/2}=60$, and include electron-ion collisions. The
following resolution parameters are used: 8 energies, 8 pitch angles,
highest energy grid point at $v=6v_a$, 6 radial grid points, 10
poloidal grid points along the orbit for circulating particles, and a
time step of $\delta t=0.01 a/c_s$.

\begin{figure}[h]
\begin{center}
\includegraphics[width=0.34\linewidth]{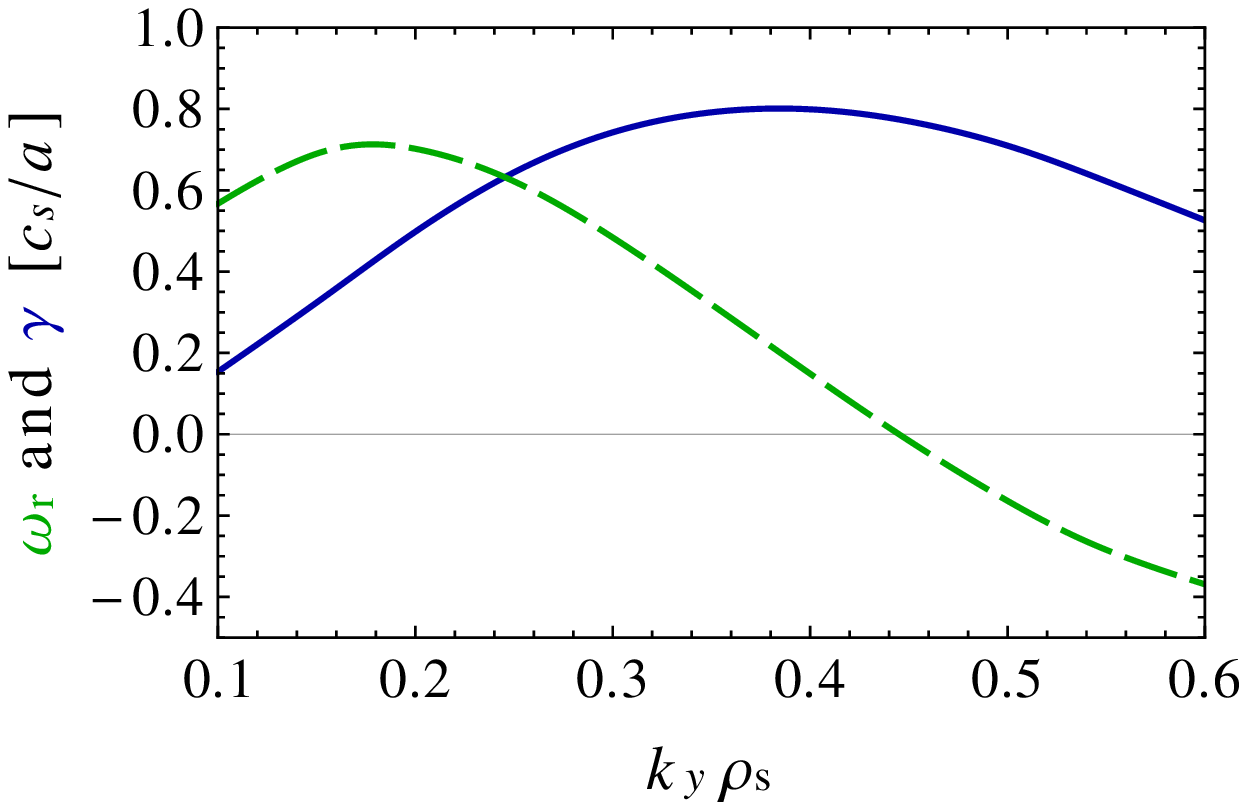}
\put(-18,80){$\rm a)$}
\put(127,80){$\rm b)$}
\includegraphics[width=0.32\linewidth]{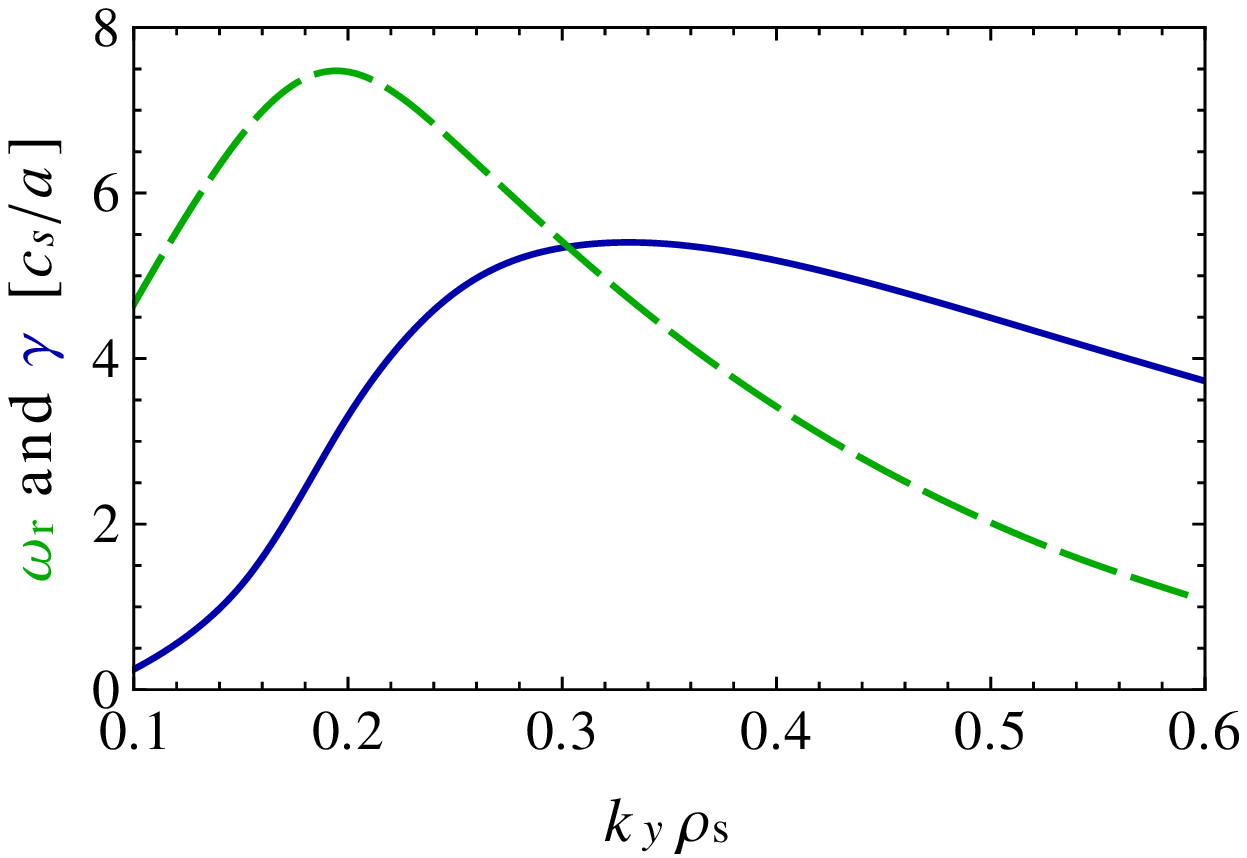}\\
\includegraphics[width=0.33\linewidth]{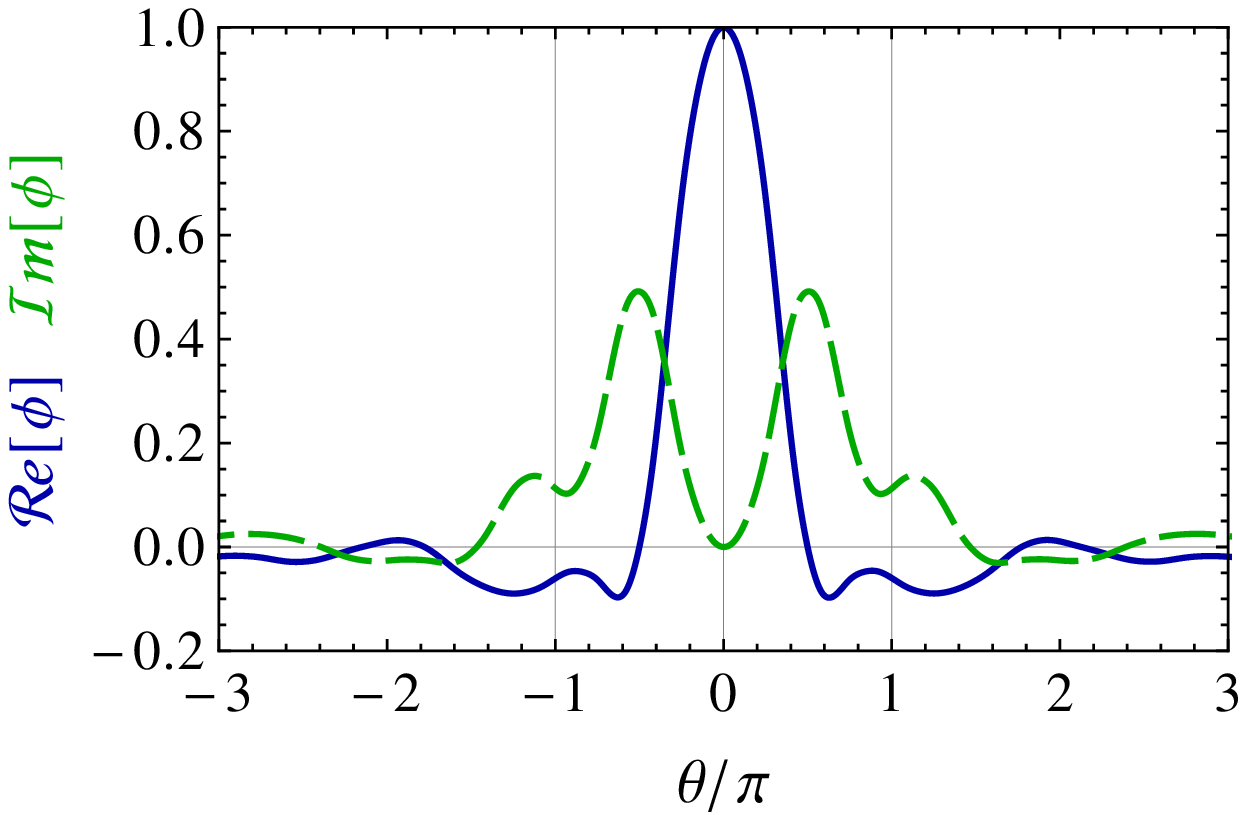}
\put(-15,80){$\rm c)$}
\put(133,80){$\rm d)$}
\includegraphics[width=0.33\linewidth]{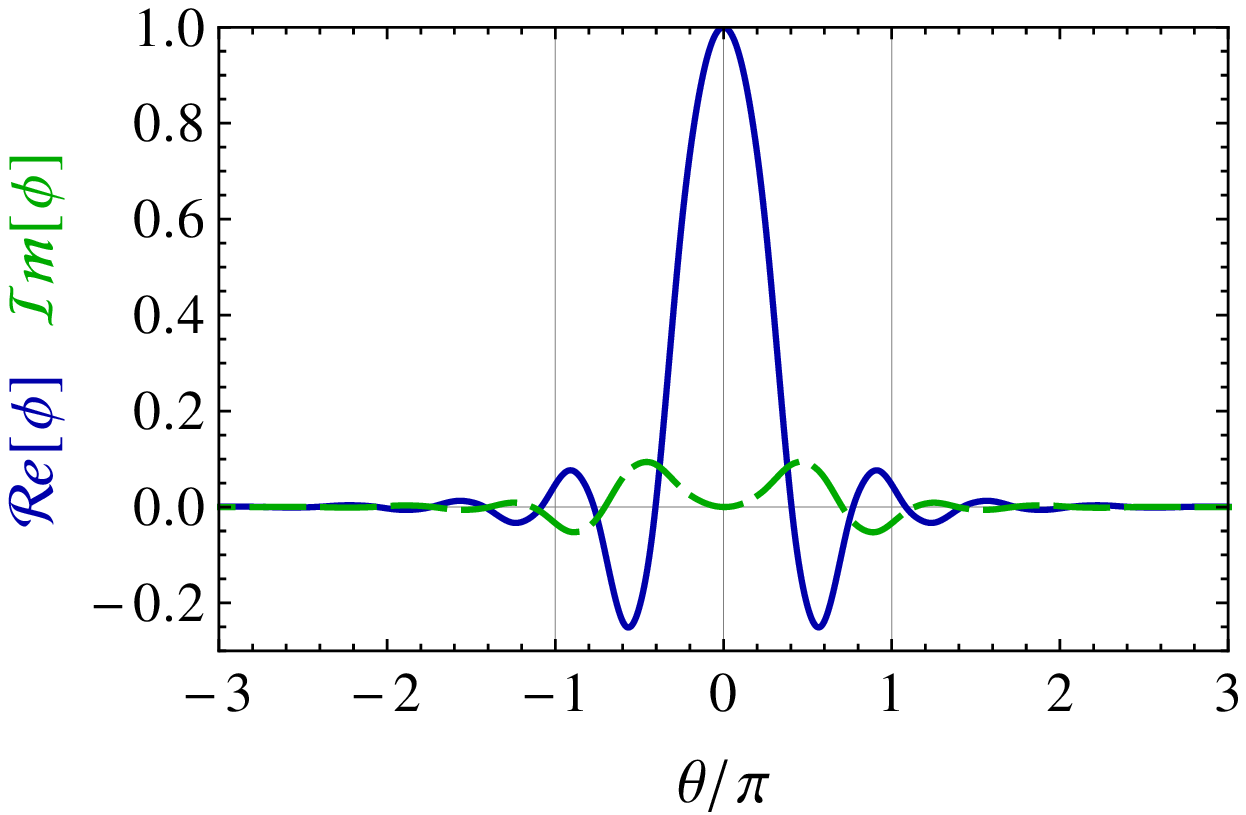}
\caption{a) and b) Binormal wave number spectra of $\omega_r$ (dashed line) and
  $\gamma$ (solid line) at $r/a=0.975$ and $r/a=0.988$,
  respectively. c) and d) Eigenfunctions $\phi$ as a function of
  $\theta$ corresponding to $k_y\rho_s=0.3$ of figures a) and b),
  respectively. Normalization: $\phi(0)=1$. Solid lines represent the
  real- and dashed lines represent the imaginary parts of
  $\phi$. Vertical grid lines mark $\theta=\{-\pi,\,0,\,\pi\}$. }
\label{gyrores}
\end{center}
\end{figure}

Figure~\ref{gyrores}a and b show the $k_y\rho_s$ spectra of real
frequencies $\omega_r$ and growth rates $\gamma$ for the cases $r/a=0.975$ and
$r/a=0.988$, respectively. Here, $\rho_s=c_s/\Omega_i$ denotes the ion
sound Larmor radius. At the inner radial position for lower wave
number the mode is propagating in the electron diamagnetic direction
(corresponding to $\omega_r>0$ according to the sign convention of
GYRO), then the propagation direction changes at wave numbers
$k_y\rho_s\approx 0.44$, as seen in Fig.~\ref{gyrores}a. We note that
a mode with qualitatively similar frequency spectrum and mode structure
exists at zero collision frequency for the same plasma parameters. The
difference is that at $\nu_{ei}=0$, $\gamma$ is somewhat higher,
$\omega_r$ is smaller, and the sign change in frequency appears closer
to $k_y\rho_s=0.3$.  At the outer radial position the mode frequency
remains positive over the plotted $k_y\rho_s$ range, and the magnitude
of the complex frequency is considerably higher than unity. The latter
is due to the very high density gradient and a corresponding high
diamagnetic frequency.  We note, without showing specific GYRO
results, that at $\nu_{ei}=0$ there exists a mode with qualitatively
similar properties in this case as well. Although one might categorize
these modes trapped electron modes based on their propagation
direction, it is important to point out that the electron collision
frequency is comparable to the typical bounce frequency of trapped
electrons in these cases.

The eigenfunctions at $k_y\rho_s=0.3$ are shown in Fig.~\ref{gyrores}c
and d. The modes are rather well localized in the extended poloidal
angle interval $\theta\in[-\pi,\,\pi]$, thus we will neglect
contributions outside this region for the analysis of the peaking
factors.

\begin{figure}[h]
\begin{center}
\includegraphics[width=0.8\linewidth]{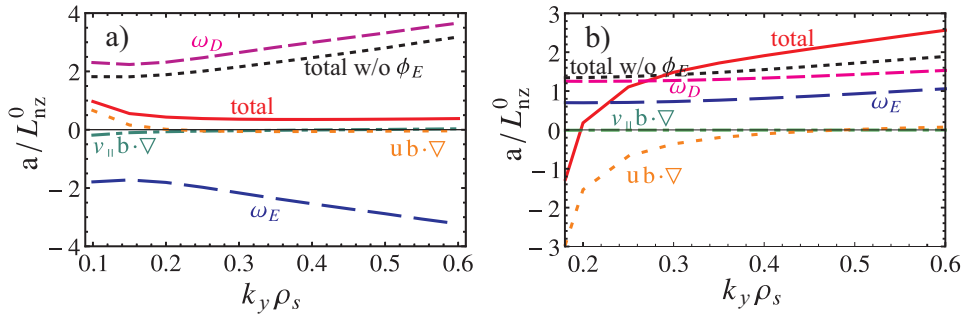}
\caption{Impurity peaking factors as a function of binormal wave
  number and contributions from the different terms of the model
  Eq.~(\ref{peak1}).  a) is at $r/a=0.975$, b) is at $r/a=0.988$. The
  peaking factor in the presence of poloidal asymmetries
  (total) is represented by the solid line, and the
  contributions from the magnetic drift ($\omega_D$), radial
  $\Ev\times\Bv$ drift ($\omega_E$), parallel compressibility
  ($\vpa\bv\cdot\nabla$) and poloidal $\Ev\times\Bv$ drift
  ($u\bv\cdot\nabla$) are represented by the dashed, the long dashed,
  the dash-dotted and the long dotted lines respectively. For
  comparison the peaking factor for the same parameters neglecting all
  electric field effects (total w/o $\phi_E$) is plotted with
  short dotted line.}
\label{aLnzfig}
\end{center}
\end{figure}

Taking the mode frequencies and eigenfunctions as input we calculate
the impurity peaking factor using the model represented by
Eq.~(\ref{peak1}). Note that while we take into account the shaping of
the flux surface in GYRO simulations, we consider a circular cross
section plasma when the peaking factors are calculated.

Figures~\ref{aLnzfig}a and b show the impurity peaking factors (total,
solid curves) as functions of the binormal wave number, $k_y\rho_s$,
for $r/a=0.975$ and $r/a=0.988$, respectively. The contributions from
the magnetic drifts ($\omega_D$, dashed), the radial $\Ev\times\Bv$
drift ($\omega_E$, long dashed), the parallel compressibility
($\vpa\bv\cdot\nabla$, dash-dotted), and the poloidal $\Ev\times\Bv$
drift ($u\bv\cdot\nabla$, long dotted) are also shown in the
figure. Since the parallel compressibility term is proportional to
$1/q^2$ which is typically a small number close to the LCFS, the
effect of parallel compressibility is negligibly small.

On the other hand, magnetic drifts are very important, especially the
radial part of the drift which is proportional to the magnetic shear
that is high at the edge. The $\omega_D$ term is always positive, and
it exhibits a weak dependence on $k_y\rho_s$ due to the weak
broadening of the eigenfunction with this parameter. The magnitude of
the contribution of $\omega_E$ to the impurity peaking is comparable
to that of the magnetic drifts, as expected since $K\sim \epsilon$. In
the in-out asymmetric case of $r/a=0.975$ the $\omega_E$ term acts to
cancel the effect of the magnetic drifts, reducing the impurity
peaking factor to almost zero. In the out-in asymmetric case of
$r/a=0.988$ the $\omega_E$ term adds to the effect of the magnetic
drifts leading to a substantial positive peaking factor at higher
values of $k_y\rho_s$. It is interesting to note that the contribution
from the magnetic drift is higher at $r/a=0.975$ than at
$r/a=0.988$. The reason is that the radial part of the magnetic drift
is $\propto s\theta\sin\theta$, thus it is weighted towards the
inboard side. In the expression for the peaking factor this function
is multiplied by the poloidally varying impurity density which, in
this case, also peaks at the inboard side. This shows that the
poloidal variation of $\mN$ in combination with the magnetic drifts do
not necessarily lead to a reduction of the peaking factor, especially
for high magnetic shear. This observation is consistent with the trend
seen in Fig.~1 of Ref.~\cite{sara1}.

For lower wave numbers $k_y\rho_s<0.3$, where usually the nonlinear
flux spectra peaks, the variation of the drift terms with $k_y$ is
very weak. However, both the $\propto U_i^2$ and the $\propto U_i$
terms of (\ref{peak1}) are inversely proportional to $k_y$, thus they
can be non-negligible for long wavelengths. At $r/a=0.975$
(Fig.~\ref{aLnzfig}a) $U_i=0.09$, and the contribution from the
$U_i^2$ term is negligibly small, the positive contribution from
$u\bv\cdot\nabla$ around $k_y\rho_s=0.1$ comes from the $U_i$ term and
is a result of the small but finite up-down asymmetry.  At $r/a=0.988$
(Fig.~\ref{aLnzfig}b) $U_i=0.68$, and although the $U_i^2$ term is
non-negligible anymore it is still the $U_i$ term that dominates the
contribution from $u\bv\cdot\nabla$.  When ${\rm
  Im}[\phi(\theta)/\phi(0)]\ll{\rm Re}[\phi(\theta)/\phi(0)]$, the
$U_i$ term is approximately proportional to $\omega_r/\gamma$ that
becomes large around $k_y\rho_i=0.1$ (see Fig.~\ref{gyrores}b). This
trend, combined with the $k_y^{-1}$ dependence of the $U_i$ term,
makes this term much bigger than all the other terms for small
$k_y\rho_s$. Accordingly, results below $k_y\rho_s=0.18$ are not shown
in Fig.~\ref{aLnzfig}b, as the perturbative treatment of the $U_i$
term breaks down.

As a comparison, we also show the impurity peaking factors for the
same plasma parameters, but without the effects of the radial and
poloidal electric fields; see the corresponding dotted curves (total
w/o $\phi_E$) in Figs.~\ref{aLnzfig}a-b. In general, this curve is
close to, but do not exactly coincide with the magnetic drift
contribution (dashed curves). Most of the difference is not due to the
parallel compressibility term, but rather the fact that in the
presence of $\phi_E$, such as for the $\omega_D$ curve, the weighting
factor $\mN$ has a poloidal variation. Notice, that the peaking factor
without poloidal asymmetries is smaller than the magnetic drift term
with poloidal asymmetries when the impurity density is peaked on the
inboard side, and \emph{vice versa}, again, in accordance with the
poloidal variation of $\mN$ and the $s\theta\sin\theta$ geometric
factor. The general observation is that the peaking factor can be
significantly modified when effects of the background electric field
are taken into account; as seen comparing the solid and the dotted
lines of Figs.~\ref{aLnzfig}a-b.

\section{Discussion and Conclusions}
\label{discussion}
We have derived an expression for the impurity density peaking factor
with corrections relevant in a steep density pedestal with subsonic
ion flows. Based on realistic edge density and temperature profiles of
a pedestal and a shaped magnetic geometry we calculate the electric
field corresponding to the neoclassical poloidal variation of the main
ion density. We use a global neoclassical solver called PERFECT that
accounts for finite orbit width effects. These effects arise in a
pedestal due to the sharp profile variations and strong radial
electric fields. Finally, we consider the effect of the poloidal and
radial electric fields on the impurity peaking due to turbulence.

We find that the poloidal location of the maximum impurity density can
vary on a short radial scale in the vicinity of the pedestal. In the
innermost radius of the simulation domain, there is an up-down
asymmetry that transits to an in-out asymmetry, which flips to an
out-in asymmetry in the strong density gradient region, then back to
an in-out asymmetry again, just inside the last closed flux surface. The
exact variation of the poloidal potential found here is not expected
to be a universal feature, it depends on the plasma parameter
profiles. However, non-negligible poloidal electric fields and their
rapid variation radially may be expected in sufficiently sharp
pedestals.

We find that neoclassical asymmetries in the edge can be strong enough
to significantly modify the turbulent impurity peaking, even though
the profiles used here for the neoclassical calculation are not
specifically chosen to maximize this effect. Due to the high safety
factor and magnetic shear close to the LCFS, the effect of parallel
compressibility is negligibly small compared to that of the magnetic-,
and the radial $\Ev\times\Bv$ drifts. The latter can reduce or enhance
the peaking factor depending on whether the poloidally varying
potential is such that the impurity density peaks in the inboard or
the outboard side. Due to the high magnetic shear close to the LCFS
the radial part of magnetic drifts is more important than the poloidal
part.

When the contribution of the $\Ev\times\Bv$ drift to the poloidal
motion of the ions is non-negligible compared to that of the parallel
streaming ($U_i\sim 1$), as in the pedestal studied here, additional
terms become important in the expression for the impurity peaking
factor (\ref{peak1}). One of them is $\propto U_i$ and it is finite
when there is some up-down variation of the long wavelength potential
$\phi_E$. The other one is formally similar to the parallel
compressibility term, but larger by a factor $2 (m_z/m_i) U_i^2$. In
our case study, even though the up-down component of the asymmetry is
small at the studied radii, the $\propto U_i$ term has a significant
contribution to the impurity peaking factor, especially at small
values of $k_y\rho_s$.

We note that we neglected impurity-ion friction effects in the
neoclassical calculation which may also modify the poloidal variation
of the impurity density.  While a sharp density pedestal can be handled by the
neoclassical simulation code used here, the code relies on the
assumption of a slow ion temperature variation. This simplifying
assumption could only be relaxed in a nonlinear, full-$f$ kinetic
code.

\begin{acknowledgement}
 This work was funded by the European Communities under Association
 Contract between EURATOM and Vetenskapsr{\aa}det (VR). IP was
 supported by the International Postdoc grant of VR. ML was supported
 by the Fusion Energy Postdoctoral Research Program administered by
 the Oak Ridge Institute for Science and Education. Fruitful
 discussions with S Moradi and valuable comments from the unknown
 referee are gratefully acknowledged.
\end{acknowledgement}

\end{document}